\documentclass[a4paper,10pt]{article}

\newcommand{\bfa}[1]{\mbox{\boldmath${#1}$}}

\newcommand{\bea}{\begin{eqnarray}}
\newcommand{\eea}{\end{eqnarray}}
\newcommand{\bnn}{\begin{eqnarray*}}
\newcommand{\enn}{\end{eqnarray*}}
\newcommand{\be}{\begin{equation}}
\newcommand{\ee}{\end{equation}}

\newcommand{\nn}{\nonumber}

\def\PACS{\par\leavevmode\hbox {\it PACS:\ }}%
\def\MSC{\par\leavevmode\hbox {\it MSC:\ }}%
\def\UK{\par\leavevmode\hbox {\it Keywords:\ }}%
\begin{document}

\title{Gravitoelectromagnetism in a complex\\Clifford algebra}

\author{S. Ulrych\\ Wehrenbachhalde 35, CH-8053 Z\"urich, Switzerland}
\date{December 10, 2005}
\maketitle

\begin{abstract}
A linear vector model of gravitation is introduced in the context of quantum physics as a
generalization of electromagnetism.
The gravitoelectromagnetic gauge symmetry corresponds
to a hyperbolic unitary extension of the usual complex phase symmetry of electromagnetism.
The reversed sign for the gravitational coupling is obtained by means
of the pseudoscalar of the underlying complex Clifford algebra.
\end{abstract}
{\scriptsize\PACS{04.50.+h; 02.10.Hh; 12.10.-g; 03.65.Pm; 11.30.Ly}
\MSC{83D05; 81V22; 11E88; 15A33; 30G35}
\UK{Gravitoelectromagnetism; Hyperbolic complex Clifford algebra; Hyperbolic numbers; Grand unification theory; Split-complex numbers}}

\section{Introduction}
The formal correspondence between Newton's law of gravitation and the electrostatic Coulomb
law has motivated many attempts to formulate gravitation in an analogy to electromagnetism.
Maxwell himself \cite{Max65} (and later Heaviside \cite{Hea93,Hea931}) turned his attention to the possibility of
formulating the theory of gravitation in a form corresponding to the electromagnetic
equations. However, the problem of the negative energy
of the gravitational field, due to the mutual attraction of material bodies, appeared
too serious to him to further follow this approach. 
Holzm\"uller \cite{Hol70} and Tisserand \cite{Tis72,Tis90} postulated that the gravitational force
of the Sun had an additional magnetic
component. This postulated gravitomagnetic component
could be adjusted to reproduce the excess perihelion precession of
Mercury. However, the solution of the relativistic Kepler problem,
without any additional interaction terms,
explains only one sixth of the discrepancy of 43 arc-seconds per century (see, e.g., Rindler \cite{Rin60}).

Some decades later Einstein predicted in his general relativity
that gravitation is mediated
by a second degree tensor 
field associated with the metric of
spacetime.
Einstein's general relativity provided an
explanation of the excess motion of Mercury's perihelion in terms of a relativistic
gravitoelectric correction to the Newtonian gravitational potential of the Sun \cite{Ein50}.
Though Einstein's theory of gravitation is substantially different to
Maxwell's vector gravitation, it could be shown that there are formal similarities.
Bel \cite{Bel58} and Penrose \cite{Pen60} have shown that the linearized Einstein
equations, perturbed about flat spacetime, can be written in a form that looks similar to
Maxwell's equations. These gravitoelectromagnetic field equations
have been derived in different forms by several authors.
For more information and further references about this topic it is referred to the 
following articles \cite{Jan92,Rug02,Tar04,Mar98,Cla00,Buc03,Mas991}.

Despite the success of Einstein's general relativity there was
again interest in a Maxwell-like vector gravitation
theory in the seventies and eighties of the last century
\cite{Maj71,Cam76, Maj82, Rai73, Rai76, Sch70, Sch71, Cav81, Sin81, Sin811}. 
The simplest way to cross over
from electromagnetism to gravity consists here in substituting for the electrical charge and
positive dielectric constant in the Maxwell equations either the imaginary gravitational
charges (Majern\'ik \cite{Maj71,Maj82}) or the negative permittivity (Brillouin \cite{Bri70}). 
Then the Coulomb and Newton laws
for two charges get the same form,
with the only difference, the opposite sign of the forces. 
The field energy density is, however, necessarily negative
(see also Richterek and Majern\'ik \cite{Ric99}).
Majern\'ik \cite{Maj72} reproduces in this context the perihelion-shift when taking into account
the self-gravity of field energy of a gravitational field.
He \cite{Maj822} has also shown that by means of 
a coupling between the gravitational and electromagnetic fields
all well-known tests of Einstein's theory of gravitation connected 
with the propagation of light in gravitational field can be correctly calculated.
The consistency of the vector theory of gravitation has been examined also
by Singh \cite{Sin82}. He modified the Hamiltonian for the
two-body interaction by a term for 
the self-interaction between particle velocity and its vector 
potential to explain the precession of the perihelion 
of a planet, the deflection of light in the gravitational field of a star, and the 
gravitational red shift, as predicted by the results of the general theory of relativity.

Motivated by these attempts to reproduce post-Newtonian gravitational effects by a vector theory of
gravitation, a Maxwell-like model for gravitoelectromagnetism is proposed in this work. The model differs not
only from Einstein's general relativity, but also from the known Maxwell-like approaches.
It is introduced in the terminology of quantum physics as a straightforward 
extension of the $U(1,\bfa{C})$ gauge symmetry of electromagnetism into the hyperbolic unitary group
$U(1,\bfa{H})$, including electromagnetism and
gravitation. 

A larger list of references regarding applications of the hyperbolic numbers \bfa{H}, 
defined as an extension of the real or the complex numbers,
has been given in \cite{Ulr05,Ulr053}. However, it should be mentioned again that they have
been applied also to general relativity, where the hyperbolic numbers are also known as
paracomplex or split-complex numbers.
The connection between differential geometry and the hypercomplex number systems 
has been shown originally by Bianchi \cite{Bia16}, 
and recently outlined again by Catoni et al. \cite{Cat05}.
Paracomplex projective models and harmonic maps were investigated by Erdem \cite{Erd83,Erd97,Erd99}.
A survey on paracomplex geometry,
para-Hermitian, and para-Kaehler manifolds has been given by Cruceanu et al. \cite{Cru95,Cru96}.
Solutions of Minkowskian sigma models generated by hyperbolic numbers were considered
by Lambert et al. \cite{Lam87,Lam88}. Zhong investigated hyperbolic complex linear symmetry groups and their local gauge
transformation actions \cite{Zho85a}. He generated
new solutions of the stationary axisymmetric Einstein equations with hyperbolic numbers \cite{Zho85}.
Furthermore, the hyperbolic complexification of Hopf algebras \cite{Zho99}.
Moffat \cite{Mof82} has interpretated the hyperbolic number as fermion number.
This interpretation has led to fundamental explanation
of stability of fermionic matter. Kunstatter et al. \cite{Kun83} investigated in this context
a generalized theory of gravitation, based on a nonsymmetric metric in a four-dimensional real manifold. 

The hyperbolic numbers are used in this work to represent the
$\bfa{R}_{3,0}$ paravector algebra, which has been introduced by
Sobczyk \cite{Sob81} for the representation of relativistic vectors. Baylis has 
shown that the theory of electrodynamics can be fully expressed in terms of
this algebra. In his textbook \cite{Bay99} a wide range of
explicit physical applications of the $\bfa{R}_{3,0}$ algebra can be found. Inserting the hyperbolic 
unit into this formalism implies, that the algebra can be complexified further
to provide the complex Clifford algebra $\bar{\bfa{C}}_{3,0}$. It has been 
proposed in \cite{Ulr052} to use this algebra to represent physical operators,
like the mass operator, in their most general form.

\section{Hyperbolic algebra}
\label{hypalg}
The commutative ring of 
hyperbolic numbers $z\in\bfa{H}$ can be defined as an extension of the complex numbers
\be
\label{beg}
z=x+iy+jv+ijw\;,\hspace{0.5cm}x,y,v,w \in\bfa{R}\;,
\ee
where the hyperbolic unit $j$ has the property $j^2=1$. 
In the terminology of Clifford algebras
the hyperbolic numbers defined in this way are represented by $\bar{\bfa{C}}_{1,0}$, i.e.,
they correspond to the universal one-dimensional complex Clifford algebra (see Porteous \cite{Por95}).

Beside the grade involution, two anti-involutions play a major role in the
description of Clifford algebras and their structure, conjugation and reversion.
Conjugation changes the sign of the complex and
the hyperbolic unit
\be
\label{conj}
\bar{z}=x-iy-jv+ijw\;.
\ee 
Reversion, denoted as $z^\dagger$, changes only the sign of the complex unit.
Anti-involutions reverse the ordering in the multiplication, e.g., $(ab)^\dagger=b^\dagger a^\dagger$.
This becomes
important when non-commuting elements of an algebra are considered.
In physics, reversion is denoted as hermitian conjugation. Note, that in \cite{Ulr052} it has been
suggested to relate hermiticity in the physical sense to the conjugation anti-involution.
With respect to conjugation the square of the 
hyperbolic number
can be calculated as
\be
\label{square}
z\bar{z}=x^2+y^2-v^2-w^2+2ij(xw-yv)\;.
\ee

The hyperbolic numbers are used to form the hyperbolic paravector algebra.
A Minkowski vector $x^\mu=(x^0,x^i)\in\bfa{R}^{\,3,1}$ is represented in terms of the
hyperbolic algebra as
\be
\label{veco}
x=x^\mu e_\mu\;.
\ee
The basis elements $e_\mu=(e_0,e_i)$ include the
unity and the Pauli algebra
multiplied by the hyperbolic
unit~$j$ 
\be
 e_\mu=(1,j\sigma_i)\;.
\ee
The only non-trivial expressions that can be generated by multiplication of the basis elements are
$j\sigma_i$, $i\sigma_i$, and $ij$. Together with the unity they form
the eight-dimensional algebra $\bfa{R}_{3,0}$.
The algebra can be complexified with either the hyperbolic or the complex unit,
which provides the additional elements $i$, $j$, $\sigma_i$, and $ij\sigma_i$.
The full structure is equivalent to the universal complex Clifford
algebra $\bar{\bfa{C}}_{3,0}$. The complexified algebra includes sixteen real dimensions. 
The pseudoscalar of the
hyperbolic algebra corresponds to
\be
ij=e_0\bar{e}_1e_2\bar{e}_3\;.
\ee

The scalar product of two Minkowski vectors
is defined as
\be
\label{scalar}
x\cdot y
=\frac{1}{2}(x\bar{y}+y\bar{x})\;.
\ee 
The basis elements of the $\bar{\bfa{C}}_{3,0}$ paravector algebra can be considered
as the basis vectors of the relativistic vector space. 
These basis elements form a non-cartesian orthogonal basis with respect to the
scalar product
\be
 e_{\mu}\cdot e_{\nu}
=g_{\mu\nu}\;,
\ee
where $g_{\mu\nu}$ is the metric tensor of the Minkowski space.

The group $SU(2,\bfa{H})$ corresponds to the spin group of the hyperbolic algebra
and its elements can
be used to express rotations and boosts of the
paravectors.
The rotation of a paravector can be represented as
\be
\label{rota}
x\rightarrow x^\prime=R x\, R^\dagger\;,
\ee
For the boosts one finds the transformation rule
\be
\label{boost}
x\rightarrow x^\prime= B x B^\dagger\;.
\ee
Rotations and boosts are given as
\be
\label{rotmat}
R=\exp{(-i\sigma_i\theta^i/2)}\;,\hspace{0.5cm}B=\exp{(j\sigma_i\xi^i/2)}\;.
\ee
The infinitesimal generators of a Lorentz
transformation can be identified as
\be
\label{gener}
J_i=\sigma_i/2\;,\hspace{0.5cm}K_i= ij\sigma_i/2\;. 
\ee
The generators satisfy the  
Lie algebra of the Lorentz
group.

Boosts are invariant under reversion
$B^\dagger = B$, whereas the conjugated
boost corresponds to the inverse
$\bar{B}=B^{-1}$. For rotations reversion and conjugation correspond both to the inverse
$R^\dagger=\bar{R}=R^{-1}$. 
This relationship indicates that in non-relativistic physics
the hermiticity of operators can be defined either
with respect to reversion or conjugation.
The effect of conjugation, reversion, and graduation on the
used hypercomplex units is displayed in Table \ref{invo}.

\begin{table}
\begin{center}
\begin{tabular}{|c|c|c|c|}
\hline
$a$ & $\bar{a}$ & $a^\dagger$ & $\hat{a}$ \\
\hline
$e_0$ & $+$ & $+$ & $+$\\
\hline
$e_i$ & $-$ & $+$ & $-$\\
\hline
$\sigma_i$ & $+$ & $+$ & $+$\\
\hline
$i$ & $-$ & $-$ & $+$\\
\hline
$j$ & $-$ & $+$ & $-$\\
\hline
\end{tabular}
\end{center}
\caption{Effect of conjugation, reversion, and graduation on the used hypercomplex units.\label{invo}}
\end{table}
Note, that graduation is an involution, which 
does not reverse the ordering in a product, i.e., $\widehat{ab}=\hat{a}\hat{b}$.
Conjugation, reversion, and graduation are related by $\bar{a}=\hat{a}^\dagger$.

This was a brief summary of the most important facts. A more detailed representation of
the hyperbolic algebra can be found in \cite{Ulr05}.

\section{Maxwell-like model of gravitation}
\label{gravi}
The electromagnetic vector potential is
attractive for unequal and repulsive for equal charges. For gravitation one
expects the potential to be attractive for equal charges and repulsive for
unequal charges. The proposal made here is to extend the $U(1,\bfa{C})$ gauge
symmetry of electromagnetism to the hyperbolic numbers. The $U(1,\bfa{H})$ phase transformation is thus 
written as
\be
C=\exp{(-i(\Lambda+ij\Lambda_g))}\;.
\ee
This global gauge transformation is unitary with respect to conjugation
\be
C\bar{C}=1\;.
\ee

The phase transformation can be extended to local gauge transformations $C(x)$.
The mass operator \cite{Ulr05} is then modified to be invariant
under these transformations.
\be
\label{basic}
M^2=(p-V(x))(\bar{p}-\bar{V}(x))\;,
\ee
where $p=i\partial^\mu e_\mu$ corresponds to the momentum operator.
The vector potential $V(x)=V^\mu(x) e_\mu$ is a combination of electromagnetic and gravitoelectromagnetic
contributions
\be
\label{field}
V(x) = eA(x)  +ijgA_g(x) \;,
\ee
where $g$ denotes the gravitoelectric charge.
Eq.~(\ref{basic}) implies a coupling between the electromagnetic and
the gravitoelectromagnetic fields.
To simplify the considerations this coupling is neglected in the following.

The mass operator is acting on the spinor field with the squared mass of the state as its eigenvalue
\be
\label{equat}
M^2\psi(x)=m^2\psi(x)\;.
\ee
The hyperbolic spinor is represented here as a  two-component column spinor $\psi^i\in \bar{\bfa{H}}^2$. 
This implies that the Pauli algebra is given in terms of the Pauli matrices.
The bar symbol indicates that the correlation, which maps the elements of the spinor to its
dual space, is represented with transposition and conjugation as given in Eq.~(\ref{conj}).
Note, that the spinor can be represented also in an algebraic form \cite{Ulr053}.

The mass equation for the electromagnetic vector potential is defined as
\be
\label{elmag}
M^2 A(x) = -J(x)\;,
\ee
with the mass operator $M^2=p\bar{p}$ and the current $J(x)=J^\mu(x)e_\mu$. 
Explicitly, this equation
can be written as \cite{Ulr05}
\bea
\label{maxwell}
M^2A&=&-\bfa{\nabla}\cdot\bfa{E}-\partial^0 C\nonumber\\
                  &&-j(\bfa{\nabla}\times\bfa{B}-
                       \partial^0\bfa{E}
                      -\bfa{\nabla}C)\nonumber\\
                  &&-i(\bfa{\nabla}\times\bfa{E}+\partial^0\bfa{B})\\
                  &&+ij\bfa{\nabla}\cdot\bfa{B}=-\rho-j\bfa{J}\nn\;,
\eea
where $C=\partial_\mu A^\mu$ disappears in the Lorentz gauge. 
Note, that the Pauli algebra is implicitly part of the three-dimensional vectors, e.g., $\bfa{E}=E^i\sigma_i$.

It is now proposed that the equations for the
gravitoelectromagnetic fields have exactly the same form as for the electromagnetic fields. 
The reversion of sign for the gravitoelectromagnetic coupling is realized in the hyperbolic algebra with the 
following mechanism. 
A Lagrangian of the form
\be
\mathcal{L}(x)=\bar{\psi}M^2\psi-m^2\bar{\psi}\psi\;
\ee
is assumed for the spinor field. Eqs.~(\ref{basic}) and (\ref{field}) then imply that the gravitoelectromagnetic current
is proportional to the gravitoelectric charge
multiplied by the pseudoscalar of the
hyperbolic algebra
\be
J \propto e\;,\hspace{0.5cm}J_g \propto ijg\;.
\ee
From Eq.~(\ref{elmag}) it follows that also the gravitoelectromagnetic vector potential
is proportional to the pseudoscalar
\be
A_g \propto ijg\;.
\ee
Reinserting this relationship into Eqs.~(\ref{basic}) and (\ref{field}) leads
to a reversion of sign for the gravitoelectromagnetic coupling by the
square of the pseudoscalar $(ij)^2= -1$. This mechanism provides
an attractive potential for equally charged particles.

Note, that the model is thus very close to the one of Majern\'ik \cite{Maj71}. Instead of current and fields
proportional to the complex unit $i$, the corresponding quantities are
related here to the pseudoscalar of the hyperbolic algebra $ij$.

\section{Post-Newtonian effects}
The presented model benefits like all Maxwell-like approaches from the close
analogy to electrodynamics and classical
field theory.
For the description of the gravitational field of big macroscropic objects
like planets and stars common classical methods can be applied.
In this context the concept of Majern\'ik is used and the coupling of the
electromagnetic charge is replaced in the formulas by the imaginary mass, here by $ijm$.

Based on this concept
the Lagrangian of a planet moving in the potential of the Sun is 
introduced in the framework of relativistic classical mechanics
as \cite{Lan62,Gol83}
\be
\label{claslag}
\mathcal{L}=-mc^2\sqrt{1-\beta^2} -ijm\phi +\frac{ijm}{c}\bfa{A}\cdot\bfa{v}\;,
\ee
where the velocity of light $c$ is displayed explicitly, $\beta=v/c$, and $m$ denotes the mass
of the planet. 
Following the approach of Singh \cite{Sin82},
the spatial vector potential is introduced as \cite{Lan62}
\be
\bfa{A}=\frac{\bfa{v}\phi}{c}\;.
\ee
The potential of the Sun is defined in complete analogy to
electromagnetism using the replacement rule for the electric charge
\be
\phi=\frac{ijM}{r}\;,
\ee
where $M$ denotes the mass of the Sun.
The hyperbolic complex contributions now drop out in Eq.~(\ref{claslag}) by $(ij)^2=-1$ and the Lagrangian becomes identical
to the Lagrangian used by Singh. 

One can then follow his calculations showing that the total energy equation of the
Sun planet system is identical to the solution obtained from Einstein's general relativity, which
is used to derive the precession of the perihelion of a planet.
Similarly, one can follow the arguments of Singh to explain the deflection of light in the
gravitational field of a star, and the gravitational redshift.

\section{Discussion}
Despite all known concerns with respect to a vector theory of gravitation
it is proposed to reconsider a Maxwell-like form of the interaction. The gravitoelectromagnetic
interaction is introduced in the context of quantum physics in analogy to electromagnetism
with the help of a hyperbolic unitary gauge symmetry. A gravitoelectric charge
is introduced, which stands in contrast to the concept of a mass charge in
the theories of Newton, Maxwell-Heaviside, and Einstein. This
gravitoelectric charge is related to a non-compact symmetry group and therefore
might appear in nature in a complete different form than the electric charge.

If the model can be further justified this would lead 
to a significant simplification of the physical theories.
The combined $U(1,\bfa{H})$ gauge symmetry of electromagnetism and gravitation is naturally
included as a substructure in the internal $SU(4,\bfa{H})$ symmetry, which is induced by
the generalized mass operator proposed in \cite{Ulr052}. This symmetry
is isomorphic to the $SU(4,\bfa{C})\times SU(4,\bfa{C})$ gauge group of the Pati-Salam model \cite{Pat73}, with the consequence that the
Pati-Salam model could be considered not only as a model for a unified
theory of the standard model, but also as a unified theory for all
known interactions.

It is assumed that the proposed model is the simplest way to introduce
gravitation into the concept of the hyperbolic $\bar{\bfa{C}}_{3,0}$ representation
of physics. Due to the success of Einstein's general relativity, 
the presented Maxwell-like model should appear as a substructure within general relativity
in the macroscopic context.

It is possible to represent general relativity in an algebraic form as an extension
of the $\bfa{R}_{1,3}$ Dirac algebra to curved spacetime. 
For a detailed discussion it is referred to Lasenby et al.
and Hestenes \cite{Las98,Dor03,Hes05}. Similar considerations could be made also for the
$\bfa{R}_{3,0}$ and $\bar{\bfa{C}}_{3,0}$ algebras, respectively.
It has been shown by Bianchi that 
the structure constants of hypercomplex numbers can be written in the same way
as the coefficients of connection \cite{Bia16}. Following the concept of Singh and with the help of an appropriate algebraic
representation of gravitation, it might be possible to derive further direct
relationships between Einstein's general relativity and the Maxwell-like approach presented in this work.

\section{Acknowledgements}
The author would like to thank Dr. Ewald Lehmann and Dr. Evi Bender for helpful discussions about gravitation.

\end{document}